# *Q*-factor mediated quasi-BIC resonances coupling in asymmetric dimer lattices

YIXIAO GAO,[1,3,4,*] LEI XU,[2] XIANG SHEN[1,3,4]

[1]*Laboratory of Infrared Materials and Devices, Research Institute of Advanced Technologies, Ningbo University, Ningbo, Zhejiang 315211, China*
[2]*Advanced Optics & Photonics Laboratory, Department of Engineering, School of Science and Technology, Nottingham Trent University, Nottingham NG11 8NS, United Kingdom*
[3]*Key Laboratory of Photoelectric Detection Materials and Devices of Zhejiang Province, Ningbo, Zhejiang 315211, China*
[4]*Engineering Research Center for Advanced Infrared Photoelectric Materials and Devices of Zhejiang Province, Ningbo University, Ningbo 315211, China*
*gaoyixiao@nbu.edu.cn*

**Abstract**: Resonance coupling in the regime of bound states in the continuum (BICs) provides an efficient method for engineering nanostructure's optical response with various lineshape while maintaining an ultra-narrow linewidth feature, where the quality factor of resonances plays a crucial role. Independent manipulation of the *Q* factors of BIC resonances enables full control of interaction behavior as well as both near- and far-field light engineering. In this paper, we harness reflection symmetry (RS) and translational symmetry (TS) protected BIC resonances supported in an asymmetric dimer lattice and investigate *Q*-factor-mediated resonance coupling behavior under controlled TS and RS perturbations. We focus on in-plane electrical dipole BIC ($ED_i$-BIC) and magnetic dipole BIC (MD-BIC) which are protected by RS, and out-of-plane electrical dipole BIC ($ED_o$-BIC) protected by TS. The coupling between $ED_i$-BIC and $ED_o$-BIC exhibits a resonance crossing behavior where the transmission spectrum at the crossing could be tuned flexibly, showing an electromagnetically induced transparency lineshape or satisfying the lattice Kerker condition with pure phase modulation capability depending on TS and RS perturbed *Q* factors. While the coupling between MD-BIC and $ED_o$-BIC shows an avoided resonance crossing behavior, where the strongly coupled resonances would lead to the formation of a Friedrich–Wintgen BICs whose spectral position could also be shifted by tuning the *Q* factors. Our results suggest an intriguing platform to explore BIC resonance interactions with independent *Q* factor manipulation capability for realizing multi-functional meta-devices.

## 1. Introduction

Optical bound states in the continuum (BICs) have attracted substantial research interests due to their intriguing capability of perfect confinement of light waves [1-4], which provide a promising scheme for constructing high-*Q* resonators for various all-dielectric structures ranging from single nanoparticles to periodic structures [2, 3, 5-7], An ideal BIC resonance in photonic systems is nonradiating state with an infinite quality factor (*Q* factor), and cannot be accessed by external incident fields for practical applications. Symmetry incompatibility and destructive interference of resonances are two main mechanisms underlying the formation of BIC. Introducing structural perturbation could open the radiation channel and transform an ideal BIC into its quasi-state with a controllable *Q* factor. For symmetry-protected BICs in metasurfaces, engineering the reflection symmetry (RS) [3, 8, 9] or translational symmetry (TS) [10-13] of a unit cell is key to constructing high-Q quasi-BIC resonances, which have been widely exploited in nonlinear harmonic generations [14, 15], nanolasers [16, 17], sensitivity-enhanced sensing [18, 19], optical modulator [20] and light-matter interactions [21, 22].

Resonance coupling offers a new degree of freedom to tailor optical responses in photonic systems, including photonic crystal slabs [23], plasmonic cavities [24, 25], whisper gallery resonators [26, 27], and cavity-coupled waveguides [28]. The coupled BIC resonances would also enrich their applications beyond their high-$Q$ feature, for example, graphene coupled with a single BIC resonance could only lead to a maximum 50% absorption [29] when the critical coupling criterion is met. Recently, Tian et al. achieved a near-unity absorption by spectrally overlapping electric-dipole BIC (ED-BIC) and magnetic dipole BIC (MD-BIC) in a lossy crystalline silicon metasurface [30], and such a mechanism could further be explored to enhance the absorption of 2D materials coupled with pure dielectric systems. While for a lossless case, the spectrally overlapped BIC resonances with matched $Q$ factors reach the lattice Kerker condition [31], forming an extreme Huygens' metasurface [32] with strongly suppressed reflection as well as highly dispersive transmissive phase modulation over a $2\pi$ range [33]. When a high-$Q$ quasi-BIC resonance couples to a broadband MD resonance, an electromagnetically induced transparency effect would take place in an all-dielectric disk metasurface [34]. Friedrich–Wintgen BICs is usually formed by destructive interferences of two resonances [17, 35], the coupling between two non-orthogonal QBIC resonances could enter strong modal coupling regime with a characteristic avoided crossing behavior as well as a vanishing linewidth in one of the splitting branches [36, 37], which could offer an interesting route for constructing Friedrich–Wintgen BICs. Nevertheless, the quality factors of the coupled BIC resonances could hardly be tuned independently in above-mentioned works, because these BICs are usually protected by the same symmetries. Any perturbation would simultaneously alter the $Q$ factors of both resonances. Realizing independently tuning the $Q$ factors of multiple BICs resonances is highly desired in order to achieve photonic devices with multi-functions.

In this paper, we investigate the coupling between BICs protected by translational symmetry (TS) and reflection symmetry (RS) in an asymmetric dielectric dimer lattice, where in-plane $ED_i$-BIC and MD-BIC protected by RS and out-of-plane $ED_o$-BIC protected by TS. RS-protected BICs are robust to TS perturbations and vice versa, and thus the quality factors of the interacting BICs protected by different symmetries could be independently controlled. We first study the coupling between two orthogonal QBICs, i.e. $ED_o$-QBIC and $ED_i$-QBIC, and the spectral response could exhibit an electromagnetically induced transparency (EIT) feature or share the same feature as the extreme Huygens metasurface, which is dependent on the relative quality factors of the two interacting QBICs. Then we study the coupling between non-orthogonal QBICs, i.e. $ED_o$-QBIC and MD-QBIC, and the spectral response demonstrates an avoided crossing feature where a Friedrich-Wintgen BIC (FW-BIC) could be formed in the lower branch of the split spectra, and the spectral position of FW-BIC could also be controlled by the quality factors of QBICs. Our findings may provide an effective method to tailor optical response through $Q$-factor mediated QBIC interaction in dielectric metasurfaces.

## 2. Structure and BIC resonance properties

Figure 1 shows the schematic of the dimer lattice investigated in this paper. The dimers are arranged in a square lattice with a period of $P$ along both $x$ and $y$ axes. As depicted in Fig. 1(b), each dimer consists of a pair of dielectric z-shaped cubes, and half of the dimer spacing is $d_1$ and the other half is $d_2$. The total length along $x$ and the width along $y$ of the z-shaped cube is $l_x$ and $l_y$, respectively. The height of the dimer is $l_z$. Without loss of generality, we consider the refractive index of the dimer lattice is 3.5, and the homogenous background has a refractive index of 1 in this paper. The lattice is excited by normally incident plane wave polarized along $y$. When $d_1 = d_2$, the dimer becomes a pair of square cubes, and if $d_1 = d_2 = (P - 2l_y)/2$, the lattice period along $y$ axis becomes $P/2$ with each unit cell containing one cube, which would be referred to as unperturbed dimer lattice as the dashed frame depicted in Fig. 1(c). In this case, the unperturbed dimer unit cell is invariant under 180 degree rotation around the $z$ axis, which corresponding to a $C_2$ operation, also has a reflection symmetry with respect to yOz plane (i.e. invariant under $\sigma_x$ operation) and xOz plane (i.e. invariant under $\sigma_y$ operation). The dimer unit cell has a $C_{2v}$ symmetry. To characterizing the degree of symmetry breaking, here we define $d_{avg} = (d_1 + d_2)/2$, and $\Delta d = d_1 - d_2$, which are related to TS and RS breaking, respectively. When $d_{avg}$ deviates from $(P - 2l_y)/2$, the TS is broken with a doubled lattice period. When $\Delta d \neq 0$, the reflection symmetry (RS) along *x* axis is broken, which indicate the unit cell cannot repeat itself after a $\sigma_x$ operation. Two symmetry-breaking mechanism is illustrated in Fig. 1(c), and two kinds of symmetries could simultaneously be broken by jointly tuning $d_1$ and $d_2$.

We focus on three kinds of BIC resonances in the unperturbed dimer lattice, as depicted in Fig. 2(a). Here we consider the dimer parameters are $l_x = 300$ nm, $l_y = 150$ nm, $l_z = 300$ nm, $d_1 = d_2 = 350$ nm and $P = 1000$ nm. BIC resonances could be classified as (i) $ED_o$-BIC: the dimer unit cell contains a pair of anti-phased out-of-plane electric dipole resonances [38]. (ii) MD-BIC: the dimer unit cell contains a pair of anti-phased magnetic dipole resonances with their dipole moments along *y* axis, which was utilized to realizing extreme Huygens' metasurface [32], and (iii) $ED_i$-BIC, the dimer unit cell contains a pair of anti-phased in-plane electric dipole resonances, which is similar to Ref. [3].

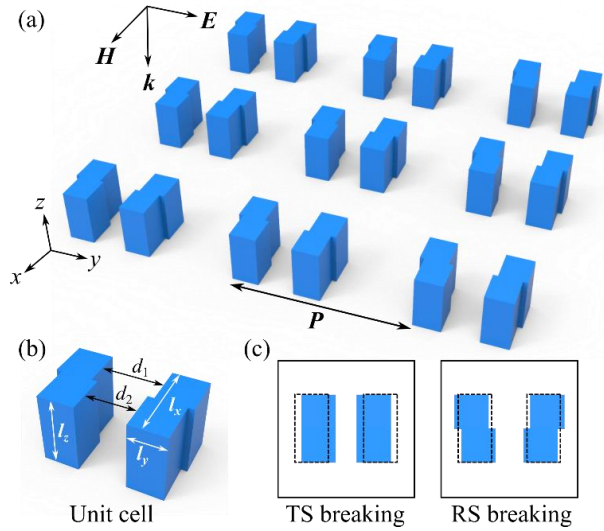

Figure 1. (a) The schematic for the dimer lattice. The dimers are arranged in a square lattice with a period of *P*. (b) The parameters for a dimer located at the center of the unit cell. (c) Perturbations of the dimer that breaks translational symmetry (left) and reflection symmetry (right). The dashed frame indicates the unperturbed dimer lattice, i.e. $d_1 = d_2 = (P - 2l_y)/2$.

We first introduce TS breaking in the unperturbed dimer lattice by shrinking $d_{avg}$ while keeping $\Delta d = 0$. Figure 2(c) shows the evolution of transmission spectra of the dimer lattice with continuously varying $d_{avg}$. When $d_{avg}$ is decreased from 350 nm to 150 nm, an emerging transmission peak at 1135.6 nm starts to blue shift accompanying a growing linewidth, indicating the $ED_o$-BIC state collapse into a quasi-state with finite *Q* factor under TS perturbation. Note that a $\pi$ phase difference could be observed in the radiating electric fields of the $ED_o$-QBIC resonance above and below the dimer lattice, as depicted in Fig. 2(b). While the MD-BIC and $ED_i$-BIC maintain their non-radiating nature upon the TS breaking, and the corresponding *Q* factors evolution under TS-only perturbation is plotted in Fig. 2(e), the calculated *Q* factors of MD-BIC and $ED_i$-BIC remains over $10^7$. Next, we introduce RS breaking by setting a nonzero value of $\Delta d$ while keeping $d_{avg} = 350$ nm. Figure 2(d) shows the transmission spectra with a varying value of $\Delta d$ from 0 nm to 100. Both MD-BIC (at 1112.2 nm) and $ED_i$-BIC (at 1142.7 nm) are sensitive to RS perturbations manifested by the increasing linewidth with $\Delta d$ increasing. However, the $ED_o$-BIC cannot be activated by the incident light throughout the parameter tuning. Fig. 2(f) shows the *Q* factors of the three BIC resonances as a function of RS-only perturbation, and the calculated *Q* factor of $ED_o$-BIC keeps over $10^7$.

Therefore, we could conclude that $ED_o$-BIC is TS protected, while MD-BIC and $ED_i$-BIC are RS protected, and the BIC protected by one symmetry is robust to the breaking of the other symmetry. It also should be pointed out that $ED_i$-QBIC has an even parity for the radiation field, and MD-QBIC and $ED_o$-QBIC show an odd parity, and the radiated electric fields toward $+z$ and $-z$ directions are in and out of phase, respectively, as depicted in Fig. 2(b).

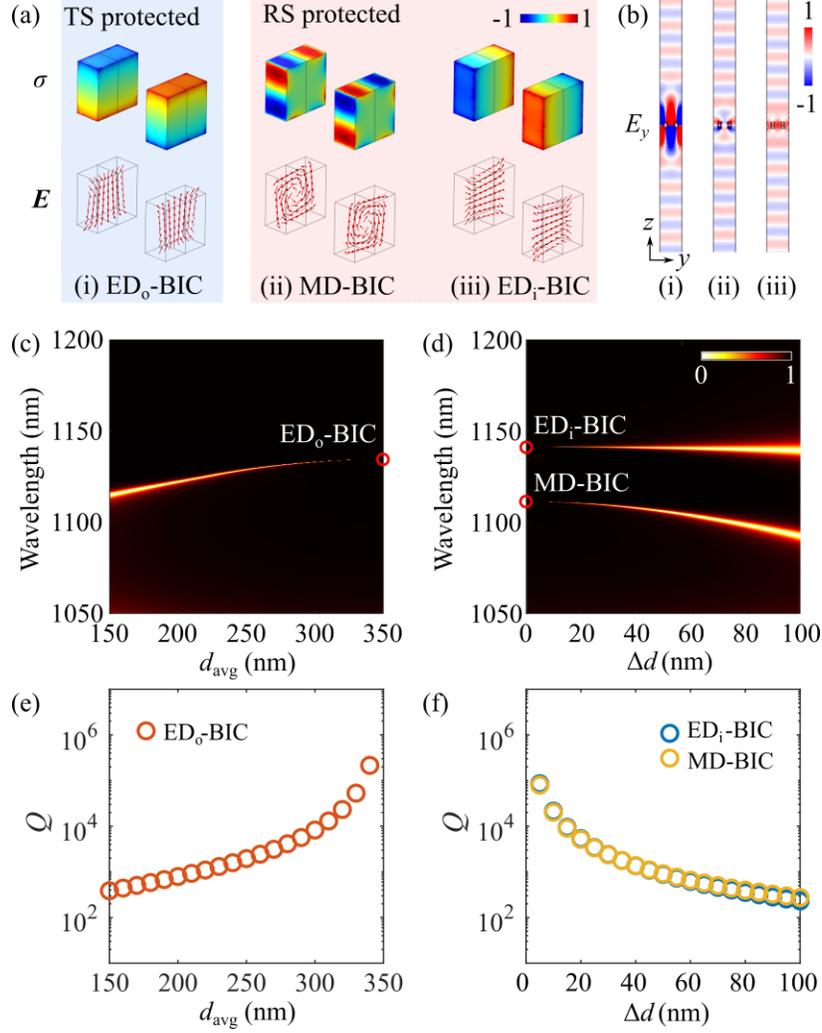

Figure 2. (a) Surface charge density (upper panel) and vectorial electric field distribution within the dimer of three BIC resonances. (b) Electric field $E_y$ in the dimer lattice at QBICs at (i) $ED_o$-QBIC, (ii) MD-QBIC and (iii) $ED_i$-BIC. The RS and TS perturbation parameters are $\Delta d = 20$ nm and $d_{avg} = 310$ nm. The color is saturated for a better illustration of the radiation field. (c) $ED_o$-BIC against TS-only perturbation with $\Delta d = 0$. (d) $ED_i$-BIC and MD-BIC against RS-only perturbation, $d_{avg} = 350$ nm. Red circles indicate the formation of BICs. The Q factors of $ED_o$-QBIC, $ED_i$-QBIC and MD-QBIC against (e) TS-only perturbation and (f) RS-only perturbation

Figure 3(a) shows the resonant wavelengths of BICs in an unperturbed dimer lattice as a function of $l_x$ calculated by eigenfrequency analysis. With the increase of $l_x$, the resonant wavelength of $ED_o$-BIC increases slower than those of $ED_i$-BIC and MD-BIC. Two intersecting points occur between $ED_o$-BIC and $ED_i$-BIC at $l_x = 290$ nm and MD-BIC and $ED_o$-BIC at $l_x = 331$ nm, respectively, indicating that we could tune $l_x$ to trigger inter-resonances coupling of two BIC states with similar resonant wavelengths. In order to activate the resonance by the external excitation, we simultaneously introduce small TS and RS perturbations to the dimer unit cell by setting $d_1$ = 280 nm and $d_2$ = 250 nm (or $\Delta d$ = 30 nm $d_{avg}$ = 265 nm). Figure 3(b) shows the transmission spectra of the perturbed dimer lattice under the varying $l_x$ from 200 nm to 400 nm. The evolution of transmission peaks is similar to the eigenfrequency calculation results plotted in Fig. 3(a). However, the transmission behaviors near the two intersections are quite different: the peaks corresponding to $ED_o$-QBIC and $ED_i$-QBIC cross each other as denoted by the yellow circle in Fig. 3(b). While MD-QBIC and $ED_o$-QBIC show an avoided crossing behavior indicating a strong coupling feature, as denoted by the white dashed circle in Fig. 3(b).

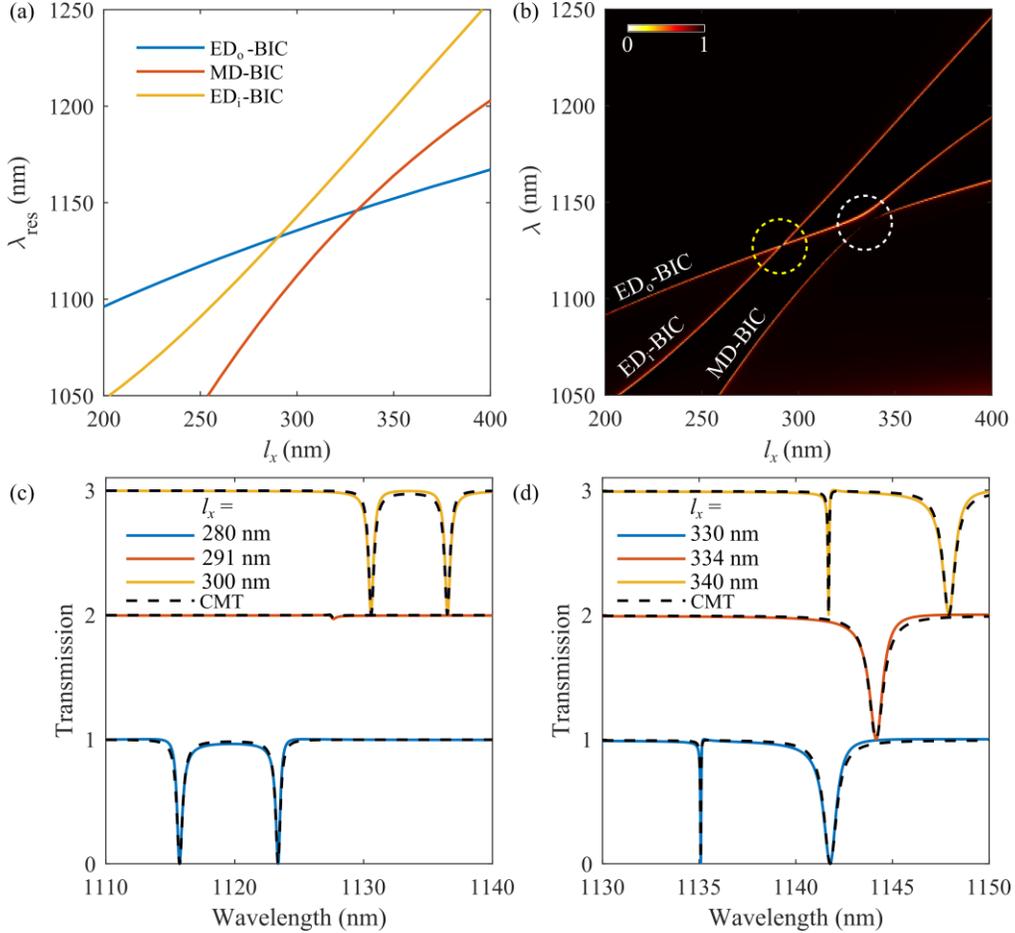

Figure 3. (a) Resonant wavelengths at $ED_o$-BIC, MD-BIC and $ED_i$-BIC as a function of $l_x$. (b) Transmission spectra of dimer lattice with varying $l_x$. Small TS and RS perturbations are introduced into the dimer unit cell by setting $d_1$ = 280 nm and $d_2$ = 250 nm. (c) Transmission spectra near the resonance crossing labeled by a yellow circle in (b). (d) Transmission spectra near the resonance anti-crossing labeled by a white circle in (b).

Before we further delve into the coupling behavior between these QBICs, we would like to reveal its underlying physics within the framework of temporal coupled mode theory [39]. The response of a perturbed dimer lattice near the modal (anti-) crossing could be modeled as a system with two resonances coupled with two ports written as [36]

$$\frac{d}{dt}\begin{bmatrix} a_1 \\ a_2 \end{bmatrix} = j \begin{bmatrix} \omega_{01} + j\gamma_1 & k + j\gamma_{12} \\ k + j\gamma_{21} & \omega_{02} + j\gamma_2 \end{bmatrix} \begin{bmatrix} a_1 \\ a_2 \end{bmatrix} + \begin{bmatrix} k_{11} & k_{12} \\ k_{21} & k_{22} \end{bmatrix} \begin{bmatrix} s_{1+} \\ s_{2+} \end{bmatrix} \qquad (1)$$

where the $[a_1, a_2]^T$ represent the time-dependent amplitudes of the QBIC resonances supported by the dimer lattice, $\omega_{01}$ ($\gamma_1$) and $\omega_{02}$ ($\gamma_2$) are resonant frequencies (decay rates) of the two QBICs, respectively. Since we do not consider the material loss of the dimer lattice, the decay rate is determined by the radiative loss, $\gamma_1$ and $\gamma_2$ are related to the quality factors ($Q_1$ and $Q_2$) of the resonances by $\gamma_1 = \omega_{01}/(2Q_1)$ and $\gamma_2 = \omega_{02}/(2Q_2)$. $k$ represents the near field coupling rate between the two modes, $\gamma_{12} = \gamma_{21}$ is the radiation coupling coefficient and $\gamma_{12} = \gamma_{21} = \sqrt{\gamma_1 \gamma_2}$. $k_{ij}$ is the coupling coefficient between the resonance $i$ and the port $j$ ($i, j = 1, 2$), and $[s_{1+}, s_{2+}]^T$ are the input wave amplitude from the port 1 and 2, respectively. The outgoing waves from the excited resonance modes are

$$\begin{bmatrix} s_{1-} \\ s_{2-} \end{bmatrix} = \begin{bmatrix} r_d & t_d \\ t_d & r_d \end{bmatrix} \begin{bmatrix} s_{1+} \\ s_{2+} \end{bmatrix} + \begin{bmatrix} d_{11} & d_{12} \\ d_{21} & d_{22} \end{bmatrix} \begin{bmatrix} a_1 \\ a_2 \end{bmatrix} \qquad (2)$$

in which $[s_{1-}, s_{2-}]^T$ are the outgoing wave amplitude at the port 1 and 2, respectively, $r_d$ and $t_d$ are the direct reflection and transmission coefficient between the ports in the absence of the resonant modes, $d_{ij}$ is the coupling coefficient between the port $j$ and the resonance $i$. The energy conservation law requires $k_{11} = k_{21} = d_{11} = d_{21} = \sqrt{\gamma_1}$ and $k_{22} = k_{12} = d_{22} = d_{12} = \sqrt{\gamma_2}$. When the input wave is incident from port 1, the transmission coefficient from port 1 to port 2 may be calculated by $t_{21} = s_{2-}/s_{1+}$.

By fitting the transmission spectra to the coupled mode theory, we could extract the coupling coefficient between the two resonances. Near the resonance crossing between $ED_o$-QBIC and $ED_i$-QBIC, we select three typical $l_x$ values equaling to 280 nm, 291 nm and 300 nm, and the corresponding transmission spectra are plotted in Fig. 3(c), and the fitting results summarized in Table 1 show that the coupling coefficients $k$ are zero for all $l_x$ values, indicating the orthogonality between $ED_o$-QBIC and $ED_i$-QBIC. The transmission dips of $ED_o$-QBIC and $ED_i$-QBIC overlap at $l_x$ = 291 nm, and a near-unity transmittance could be observed in the crossing region, which is related to the concept of extreme Huygens metasurface [32] where two QBIC resonances with opposite parity of the radiation field, as depicted in Fig. 2(b), largely cancels the back reflection. While for the intersection between MD-QBIC and $ED_o$-QBIC, the coupling coefficient $k$ becomes nonzero and is larger than both $\gamma_1$ and $\gamma_2$, as summarized in Table 2, indicating the interaction between MD-QBIC and $ED_o$-QBIC enters the strong mode coupling regime. Figure 3(d) shows the transmission spectra near the avoided crossing at $l_x$ equal to 330 nm, 334 nm and 340 nm. We could find that the linewidth at the lower branch is narrower than that of the upper branch, and a vanishing linewidth of the lower branch at $l_x$ = 334 nm actually indicates the formation of Friedrich–Wintgen BIC [1]. The quality factors and coupling coefficients play an important role in the QBICs coupling process, which will be further discussed in following sections.

Table 1. Fitting parameters for the crossing between $ED_o$-BIC and $ED_i$-BIC

| $l_x$ (nm) | $\lambda_{01}$ (nm) | $\lambda_{02}$ (nm) | $Q_1$ | $Q_2$ | $k$ (× 2π THz) |
|---|---|---|---|---|---|
| 280 | 1130.55 | 1136.55 | 2500 | 2200 | 0 |
| 291 | 1127.59 | 1127.55 | 2200 | 2500 | 0 |
| 300 | 1115.72 | 1123.37 | 2200 | 2500 | 0 |

Table 2. Fitting parameters for the anticrossing between MD-QBIC and $ED_o$-QBIC

| $l_x$ (nm) | $\lambda_{01}$ (nm) | $\lambda_{02}$ (nm) | $Q_1$ | $Q_2$ | $k$ (× 2π THz) |
|---|---|---|---|---|---|
| 330 | 1140.85 | 1136.00 | 2200 | 3700 | 0.532 |
| 334 | 1141.90 | 1140.70 | 2320 | 3620 | 0.640 |
| 340 | 1142.70 | 1146.91 | 3500 | 2500 | 0.522 |

## 3. Interaction of ED$_o$-QBIC and ED$_i$-QBIC resonances

We first study the resonance coupling between ED$_o$-QBIC and ED$_i$-QBIC. In order to access these resonances by free-space radiation, we simultaneously deviate $d_{avg}$ and $\Delta d$ from BIC state of the dimer lattice, and in such cases, both BIC resonances work in their quasi-states. Figure 4 shows the transmission spectra under different spacing parameters $d_i$ as a function of $l_x$. Since the two modes cannot be coupled with each other, resonance crossing behavior could be observed in all cases, while the linewidths of each resonance have an important influence on the spectral behavior of the resonance crossing.

In Fig. 4(a), we set $\Delta d = 20$ nm and $d_{avg} = 250$ nm to introduce a relatively small perturbation breaking the RS, and ED$_i$-QBIC has a narrower linewidth than the ED$_o$-QBIC. For example, at $l_x = 300$ nm, the $Q$ factors of ED$_i$-QBIC and ED$_o$-QBIC are 4599 and 1861, respectively. We extract the transmission spectra in the high transmission region at the crossing region when $l_x = 291$ nm, as denoted by the blue dashed line in Fig. 4(a), which is also plotted in Fig. 4(d), and a drop in the transmittance near 1125 nm could be observed with a minimum value of 0.76. Near the transmission dip, a $2\pi$ phase shift could also be observed due to the resonances overlapping, as depicted in Fig. 4(d).

If $\Delta d$ is increased to 40 nm while keeping $d_{avg} = 250$ nm, the linewidth of ED$_i$-QBIC becomes wider and similar to that of ED$_o$-QBIC, e.g. the $Q$ factors of the ED$_i$-QBIC and ED$_o$-QBIC are 1249 and 1501 at $l_x = 300$ nm, respectively. We plot the transmission spectra at $l_x = 292$ nm in Fig. 4(d), as denoted by the red dashed line in Fig. 4(b), and the spectra becomes nearly flat with a minimum transmittance of 0.96 and a $2\pi$ phase shift also occurs near the overlapped resonant wavelength, indicating the formation of a highly-transmissive extreme Huygens metasurface, with the capabilities of group velocity control and pure phase modulation [32, 33].

When $\Delta d$ is further increased to 60 nm, the linewidth of ED$_i$-QBIC is wider than ED$_o$-QBIC, e.g. the $Q$ factors of the ED$_i$-QBIC and ED$_o$-QBIC are 524 and 1120 at $l_x = 300$ nm, respectively. At the crossing region with $l_x = 293$ nm, a low transmission dip re-emerges in the transmission spectra with a minimum transmittance of 0.73. If we keep the large RS perturbation with $\Delta d = 60$ nm, while reducing the TS perturbation by setting $d_{avg} = 325$ nm, the linewidth of ED$_i$-QBIC would be much wider than ED$_o$-QBIC, as depicted in Fig. 4(c). At the resonance crossing, for example, at $l_x = 291.6$ nm, as denoted by the yellow dashed line in Fig. 4(c), the transmission spectrum demonstrates an electromagnetically induced transparency (EIT) feature [34], i.e. a narrow band of high transmittance emerges inside the broad stop band, as depicted by the yellow curve in Fig. 4(d).

Comparing the transmission spectra of the dimer lattice with different TS and RS perturbations presented in Fig. 4(d), we could find that, in order to achieve an extreme Huygens' metasurface with high transmission efficiency, the two orthogonal overlapped QBIC resonances with opposite radiation field parity should have similar linewidths to ensure a nearly flat transmission spectrum, which is highly important for pure phase modulation[20, 33, 40]. Also, the two spectrally overlapped QBICs with large $Q$ factors could exhibit a EIT-like spectrum, indicating the lineshape of the interacting QBIC resonances could be flexibly tuned by their $Q$ factors through controlling the symmetry perturbation parameters. In addition, if the material loss is introduced in the lattice [30], tuning the $Q$ factor could reach degenerate critical coupling condition [23] which could induce 100% absorption, which could be further applied to enhancing the absorption in 2D materials [29].

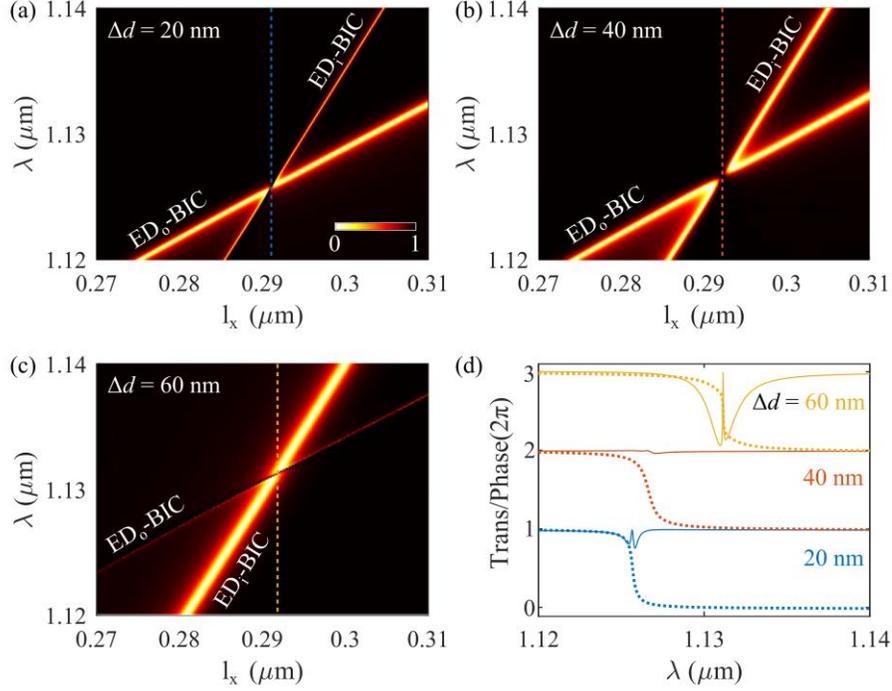

Figure 4. Resonance crossing between $ED_i$-QBIC and $ED_o$-QBIC under different perturbations (a) $\Delta d = 20$ nm and $d_{avg} = 250$ nm. (b) $\Delta d = 40$ nm and $d_{avg} = 250$ nm. (c) $\Delta d = 40$ nm and $d_{avg} = 325$ nm. (d) Transmission (solid) and phase shift (dashed) with $l_x = 291$ nm and $\Delta d = 20$ nm (blue), $l_x = 292$ nm and $\Delta d = 40$ nm (red), $l_x = 291.6$ nm and $\Delta d = 60$ nm (yellow).

## 4. Interaction of $ED_o$-QBIC and MD-QBIC resonances

Next, we investigate the interaction between $ED_o$-QBIC and MD-QBIC with the same phase symmetry of the radiation field. Figure 5(a) shows the transmission spectra of the dimer lattice with $l_x$ varying from 320 nm to 360 nm. Here, we introduce a RS perturbation with $\Delta d = 10$ nm and a TS perturbation with $d_{avg} = 250$ nm to the dimer lattice where MD-QBIC have in a higher $Q$ factor compared with $ED_o$-QBIC. With the growing $l_x$, the resonant wavelength of MD-QBIC is increasing faster than that of $ED_o$-QBIC. At $l_x \sim 334$ nm, the transmission spectra exhibit an avoided crossing feature with a spectral splitting of 2.3 nm at $l_x = 334$ nm, indicating a strong coupling occurred between two resonances.

To gain a deeper insight into the strong coupling of QBICs resonances, we perform a multipolar analysis [11, 41] on the field profile in one cube of the dimer. We use the Cartesian multipolar decomposition to identify the multipole components of resonance field in the cube, and induced multipole moments are calculated relying on the equations presented in Ref. [11]. Note that group theory approach could also provide an efficient method to analyze the formation of symmetry-protected BICs [42, 43]. Figure 5(b) depicts six exemplary vectorial electric field distributions in one cube labelled by different colored dots denoting the spectral position in Fig. 5(a). We note that the other cube in the dimer has an opposite field profile as depicted in Fig. 2(a). When the resonant wavelengths of $ED_o$-QBIC and MD-QBIC are well-separated and far from the avoided crossing region, for example, at $l_x = 320$ nm, the resonance field in the upper branch mainly oscillates along the $z$ axis, and have an out-of-plane electric dipolar moment $p_z$, while the resonance field in the lower branch mainly contains a magnetic dipolar moment $m_y$, which could also be observed in the vortex field distribution, as depicted in Fig. 5(b). With $l_x$ increasing, the resonant wavelengths of $ED_o$-QBIC and MD-QBIC are approaching each other, while the resonance strong occurs and two transmission dips repel each other. The resonance fields of two QBICs hybridize with each other, at $l_x = 334$ nm, both resonance field show a vortex field distribution with its center deviated from the cube center, which results in each cube simultaneously contains an electric dipole moment aligning with $z$ axis and magnetic dipole moment aligning with $y$ axis, which is confirmed by a multipolar expansion result presented in the Fig. 5(b). When $l_x$ is further increased to 345 nm, the resonant wavelengths of $ED_o$-QBIC and MD-QBIC are separated again, and away from the strong coupling region, the resonance field in the upper branch return to a vortex distribution with a dominant MD component $m_y$, while the lower branch has a dominant ED component $p_z$, and a slightly bended electric field on one side of the cube could be observed, indicating $ED_o$-QBIC is still slightly hybridized MD-QBIC.

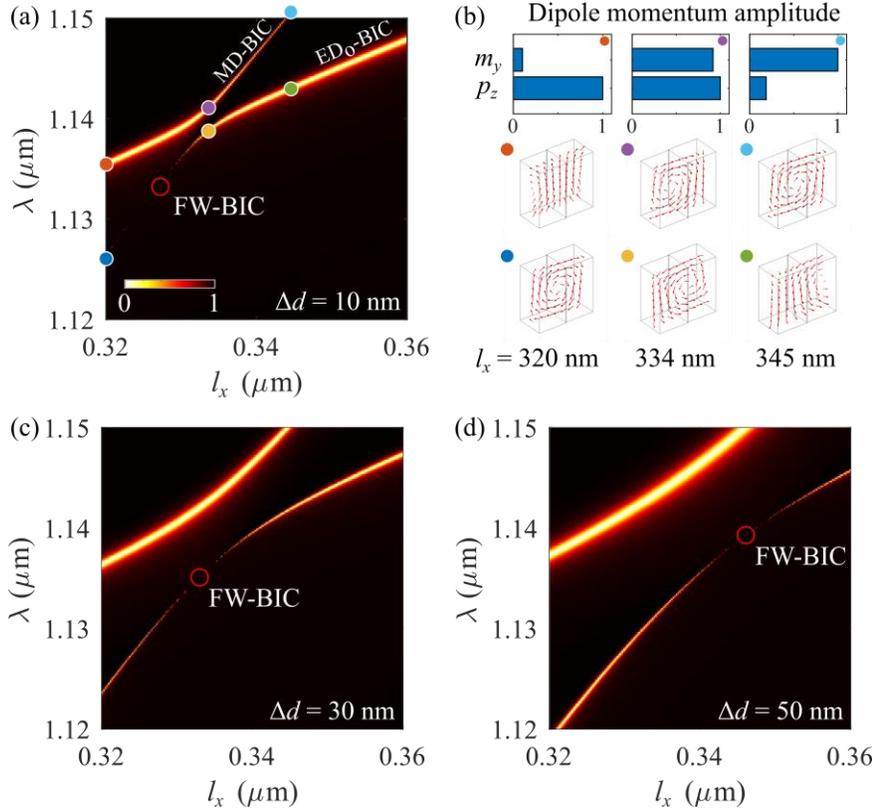

Figure 5. Resonance avoided crossing between $ED_i$-QBIC and MD-QBIC under continuous $l_x$ tuning. (a) Transmission spectra with $\Delta d = 10$ nm. (b) Normalized amplitudes of magnetic dipole $m_y$ and electric dipole $p_z$ of the upper branch QBIC resonance at the colored points in (a), and the vectorial electric field distribution in one cube of the perturbed dimer at the colored points in (a). Transmission spectra with (c) $\Delta d = 30$ nm, and (d) $\Delta d = 50$ nm. The red circles indicate the spectral position of FW-BIC.

When the RS perturbation is increased, for example, $\Delta d = 30$ nm, the transmission spectra with varying $l_x$ is plotted in Fig. 5(c). The resonance splitting is 6.4 nm at $l_x = 336.7$ nm, which is larger compared with that of $\Delta d = 10$ nm. If we further increase $\Delta d$ to 50 nm, the resonance splitting is increased to 10.7 nm at $l_x = 340.7$ nm. When we fix $\Delta d = 10$ nm while changing $d_{avg}$ to tune TS perturbation, we also found the spectral splitting will be larger in a smaller $d_{avg}$ value. Therefore, we could conclude that the coupling strength between $ED_o$-QBIC and MD-QBIC is proportional to perturbation strength.

Near the strong coupling region, a vanishing linewidth in the lower branch could also be observed, which is denoted by the red circles in Fig. 5(a, c, and d). We notice that this phenomenon is related to Friedrich-Wintgen BIC (FW-BIC) [1], and occurs when the criterion $\kappa(\gamma_1 - \gamma_2) = \sqrt{\gamma_1\gamma_2}(\omega_1 - \omega_2)$ is met where destructive interference between two coupled resonances leads to total suppression of radiation. By tuning the perturbation strength, the spectral position of FW-BIC would be shifted along the lower branch, due to the joint interaction between resonance parameters of two coupled QBICs.

## 5. The influence of a substrate and material loss

A substrate would have an important influence on the BIC resonances, for example, a substrate could transform a BIC into a leaky resonance due to the opening diffraction channels [44]. If the resonant wavelength of dimer lattice is larger than the product of dimer lattice period and refractive index of substrate, a BIC resonance could still be maintained. Here, we consider a dimer lattice sitting on a silica substrate with a refractive index of 1.45. We set the parameters of the dimer lattice as $l_y = 200$ nm, $l_z = 350$ nm, $\Delta d = 30$ nm and $d_{avg} = 265$ nm, and the $l_x$ varies from 350 nm to 700 nm. In such case, both the RS and TS are broken, and we excited the dimer lattice with a substrate with a normally incident y-polarized plane wave. The evolution of the transmission spectra with varying $l_x$ is plotted in Fig. 6(a), which is similar to Fig. 3(b). However, the main difference is denoted by the dashed box, where the $ED_i$-QBIC and $ED_o$-QBIC intersects, and the enlarged view of the dashed box is plotted in Fig. 6(b). The substrate breaks the out-of-plane symmetry possessed by the dimer lattice in a homogenous medium, leading to a coupling of the two interacting QBICs manifested by the splitting peaks. If the coupling strength is strong enough, an avoided crossing behavior could also be anticipated [28].

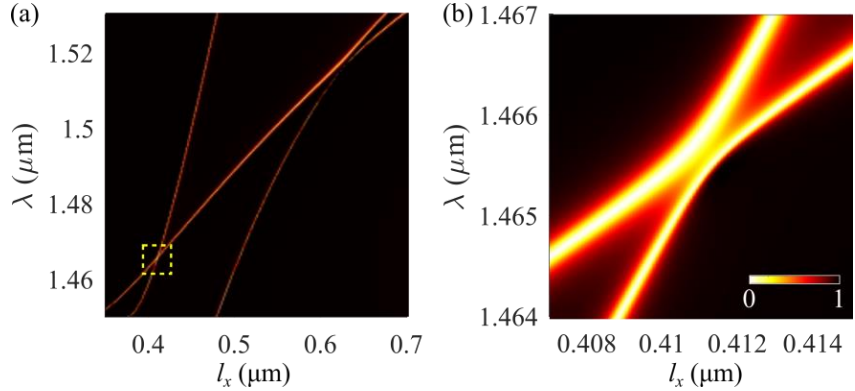

Figure 6. (a) Transmission spectra of dimer lattice on a substrate with varying $l_x$. (b) Enlarged view of the dashed box region in (a).

Next, we discuss the influence of material loss on the coupling behavior. We assume the dimer lattice has an extinction coefficient $k = 0.001$ and 0.002, while the refractive index keeps at 3.5, and other parameters are set as $l_y = 150$ nm, $l_z = 300$ nm, $\Delta d = 30$ nm and $d_{avg} = 265$ nm, and the dimer lattice is excited by a y-polarized, normally incident plane wave. The transmission spectra of a dimer lattice with and without material loss as a function of $l_x$ are plotted in Fig. 7. The resonant wavelength evolutions are similar for both the loss and lossless cases, and the linewidth for three QBIC resonances increase with a larger $k$ value as expected. The FW-BIC would also occur for the lossy case. While the material loss will decrease the transmittance (~ 0.1 when $k = 0.002$) at the crossing between $ED_i$-QBIC and $ED_o$-QBIC, which would largely deteriorate transmission efficiency of an extreme Huygens metasurface.

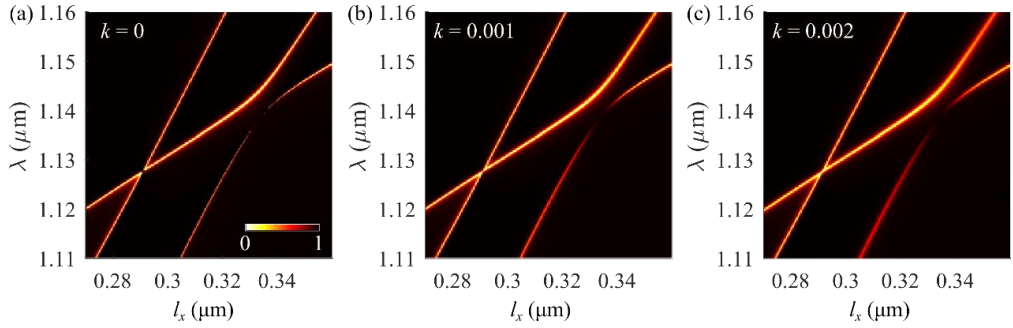

Figure 7. The transmission spectra of the dimer lattice as a function of lx with an extinction coefficient (a) k = 0, (b) k = 0.001, and (c) k = 0.002.

## 6. Conclusion

In summary, we have investigated the coupling of QBIC resonances in asymmetric dimer lattices, where we mainly focus on three typical BIC resonances, i.e. $ED_i$-BIC and MD-BIC protected by RS and $ED_o$-BIC protected by TS. The $Q$ factors of the QBIC resonance could be independently controlled by perturbing the corresponding symmetry protecting the BIC, which provides an effective way to explore the $Q$-factor-mediated resonance coupling behavior between two QBICs. The coupling between $ED_o$-QBIC and $ED_i$-QBIC, which have opposite phase symmetries of far-field radiations, demonstrates a resonance crossing behavior with a continuously dimer length $l_x$ tuning, and the transmission spectrum at the crossing could be tuned to share the same features from an extreme Huygens metasurface to an electromagnetically induced transparency effect, by adjusting the $Q$ factors of $ED_o$-QBIC and $ED_i$-QBIC through TS or RS perturbation. For the coupling between $ED_o$-QBIC and MD-QBIC with the same radiation phase symmetries, a resonance avoided crossing could be observed indicating a strong modal coupling occurs, and a hybrid ED and MD resonance is presented in the avoided crossing region revealed by a multipolar expansion analysis. The coupling strength is also proportional to the strength of the symmetry perturbation. A FW-BIC could be formed in the lower branch of split spectra due to destructive interference between $ED_o$-QBIC and MD-QBIC, whose spectral position could be shifted by controlling the $Q$ factors of two coupling QBIC resonances. Our results could provide an ideal platform for studying various resonance coupling phenomena as well as tailoring the spectral lineshape of lattice resonators, which could find important applications including nanolasing, sensing, modulating etc.

### Funding

National Natural Science Foundation of China (Grant No. 62105172), Zhejiang Provincial Natural Science Foundation of China under Grant No. LQ21F050004, Ningbo Natural Science Foundation (Grant No. 202003N4102), K. C. Wong Magna Fund in Ningbo University

### Disclosures

The authors declare no conflicts of interest.

### Data availability

Data underlying the results presented in this paper are not publicly available at this time but may be obtained from the authors upon reasonable request.